\documentclass{eptcs}
\providecommand{\event}{PLACES 2012} 

\providecommand{\urlalt}[2]{\href{#1}{#2}}
\providecommand{\doi}[1]{doi:\urlalt{http://dx.doi.org/#1}{#1}}

\usepackage{amsmath}
\usepackage{amssymb}
\usepackage{alltt}
\usepackage{mathtools}

\usepackage{tikz}
\usetikzlibrary{decorations.pathreplacing}
\usetikzlibrary{shapes}

\newcommand{\coord}[1]{\tikz[remember picture] \node[coordinate,yshift=0.8em,xshift=0.25em] (#1) {};}
\newcommand{\drawbox}[3]{\tikz[remember picture, overlay] \draw[thick] (#1) rectangle (#2);%
                         \tikz[remember picture, overlay] \node at (#2) [fill=black, text=white, above left] {\Large{#3}};}

\begin{document}

\title{Mapping the Join Calculus to Heterogeneous Hardware}
\def\titlerunning{Mapping the Join Calculus to Heterogeneous Hardware}




\author{
    Peter Calvert
    \institute{Computer Laboratory\\ University of Cambridge}
    \email{Peter.Calvert@cl.cam.ac.uk}
\and
    Alan Mycroft
    \institute{Computer Laboratory\\ University of Cambridge}
    \email{Alan.Mycroft@cl.cam.ac.uk}
}

\def\authorrunning{Calvert and Mycroft}

\maketitle
\begin{abstract}
As modern architectures introduce additional heterogeneity and parallelism, we
look for ways to deal with this that do not involve specialising software to
every platform. In this paper, we take the Join Calculus, an elegant model for
concurrent computation, and show how it can be mapped to an architecture by
a Cartesian-product-style construction, thereby making use of the calculus'
inherent non-determinism to encode placement choices. This unifies the concepts
of placement and scheduling into a single task.
\end{abstract}

\section{Introduction}\label{sec:introduction}
The {\em Join Calculus} was introduced as a model of concurrent and distributed computation \cite{Fournet1996}.
Its elegant primitives have since formed the basis of many concurrency extensions to existing languages---both functional \cite{Conchon1999,Odersky2000} and imperative \cite{Benton2004,vonItzstein2005}---and also of libraries \cite{Russo2007}.
More recently, there has also been work showing that a careful implementation can match, and even exceed, the performance of more conventional primitives \cite{Turon2011}.

However, most of the later work has considered the model in the context of shared-memory multi-core systems.
We argue that the original Join Calculus assumption of {\em distributed computation} with disjoint local memories lends itself better to an automatic approach.
This paper adapts our previous work on Petri-nets \cite{Calvert2011} to show how the non-determinism of the Join Calculus can express placement choices when mapping programs to heterogeneous systems; both data movement between cores and local computation are seen as scheduling choices.
Thus we argue that a JVM-style bytecode with Join Calculus primitives can form a universal intermediate representation, that not only models existing concurrency primitives, but also adapts to different architectures at load-time.

This paper introduces our construction by considering the following Join
Calculus program that sorts an array of integers using a {\em merge-sort}-like
algorithm. There is clearly scope for parallelising both the \texttt{split}
and \texttt{merge} steps---although this may require moving data to another
memory.
\begin{alltt}\small
  def sort(numbers, k) =
    let N = numbers.length in
    def split(a) =
          let n = a.length
          in  if n == 1 then merge(a)
                        else split(a[0..(n/2)-1]) & split(a[(n/2)..(n-1)])
        merge(a) & merge(b) =
          if a.length + b.length == N then do_merge(a, b, k)
                                      else do_merge(a, b, merge)
    in split(numbers)
\end{alltt}
where \texttt{do\_merge} is a functional-style procedure that merges the sorted
arrays \texttt{a} and \texttt{b} into a new sorted array that is passed to its
continuation (\texttt{k} or \texttt{merge}). We assume moderate familiarity with
Join Calculus primitives. In particular, we will:
\begin{itemize}
  \item Restrict the Join Calculus to make all data usage explicit, showing that
        existing programs can be desugared into this form
        (Section~\ref{sec:non-nested}).
  \item Briefly show how our existing work manifests itself in the Join Calculus
        (Section~\ref{sec:construction}).
  \item Introduce {\em workers} to the Join Calculus semantics as a substitute
        for the {\em resource constraints} in the Petri-net version
        (Section~\ref{sec:workers}).
\end{itemize}
We offer a discussion of the scheduling issues and how we believe these to be
tractable in Section~\ref{sec:scheduling}, before concluding in
Section~\ref{sec:conclusion}.

\section{The Non-Nested Join Calculus}\label{sec:non-nested}
As in our previous work, our construction introduces explicit data
transfer transitions.  For these to cover all required data transfers, we
disallow references to free variables which may be located on other
processors---i.e. values that are not received as part of the transition's
left-hand-side join pattern.  Unfortunately, nested Join Calculus definitions
capture values in this way.  In our running example, observe that both
\texttt{N} and \texttt{k} are used implicitly by \texttt{merge}.

Our formulation of the Join Calculus therefore forbids the nesting of
definitions. Instead, programs consist of a list of definitions. This
necessitates a special type of signal, {\em constructors}, that are used to
create and initialise a join definition. A new version of our program is shown
in box ``A'' of Figure~\ref{fig:code}.
\begin{figure}
\begin{alltt}\small
definition \{
 \coord{tlA} .ctor sort_x(numbers, k) \{                \coord{tlB} .ctor sort_y(numbers, k) \{
    split_x(numbers);                          split_y(numbers);
    info_x(numbers.length, k);                 info_y(numbers.length, k);
  \}                                          \}
  
  info_x(N, k) \{ info_x(N,k); info_x(N,k); \} info_y(N, k) \{ info_y(N,k); info_y(N,k); \}
  
  split_x(a) \{                               split_y(a) \{
    let n = length(a);                         let n = length(a);
    
    if(n == 1) \{ merge_x(a);               \}   if(n == 1) \{ merge_y(a);               \}
    else       \{ split_x(a[0..(n/2)-1]);       else       \{ split_y(a[0..(n/2)-1]);
                 split_x(a[(n/2)..(n-1)]); \}                split_y(a[(n/2)..(n-1)]); \}
  \}                                          \}
  
  merge_x(a) & merge_x(b) & info_x(N, k) \{   merge_y(a) & merge_y(b) & info_y(N, k) \{
    info_x(N, k);                              info_y(N, k);
    if(a.length + b.length == N)               if(a.length + b.length == N)
         do_merge(a, b, k);                         do_merge(a, b, k);
    else do_merge(a, b, merge_x);              else do_merge(a, b, merge_y);
  \}                                          \}
                                            \coord{brA}                                           \coord{brB}
 \coord{tlC} split_x(a)   \{ split_y(a);          \}      split_y(a)   \{ split_x(a);          \}
  merge_x(a)   \{ merge_y(a);          \}      merge_y(a)   \{ merge_x(a);          \}
  info_x(N, k) \{ info_y(N, "k on y"); \}      info_y(N, k) \{ info_x(N, "k on x"); \}
\}                                                                                      \coord{brC}\drawbox{tlA}{brA}{A}\drawbox{tlB}{brB}{B}\drawbox{tlC}{brC}{C}\end{alltt}
  \caption{Mapped version of {\em merge-sort} for a dual-processor system}\label{fig:code}
\end{figure}

However, despite this restriction, nested definitions can easily be
encoded by a process similar to both {\em lambda-lifting} and Java's {\em
inner-classes}.  In particular, any program similar to:
\begin{alltt}\small
  a(x,k) \{
    definition \{ .ctor Nested() \{ k(f);     \}
                 f(m)           \{ m(x * 2); \} \}
    construct Nested();
  \}
\end{alltt}
can be rewritten in a similar-style to:
\begin{alltt}\small
  definition \{ .ctor UnNested(x,k) \{ temp(x); k(f);     \}
               temp(x)             \{ temp(x); temp(x);  \}
               f(m) & temp(x)      \{ temp(x); m(x * 2); \} \}
  a(x,k)     \{ construct UnNested(x,k); \}
\end{alltt}
Unfortunately, the extra signal would cause serialisation of many transitions
within the definition. This is resolved by the {\em duplication} transition that
allows us to create as many copies of the \texttt{x} message as we require. We
rely on the scheduler not to perform excessive duplications. We might also be
able to optimise this `peek' behaviour in our compiler.

As we will later build on our previous work involving Petri-nets
\cite{Calvert2011}, it is worth highlighting Odersky's discussion
\cite{Odersky2000} on the correspondence between the Join Calculus and
(coloured) Petri-nets.  Just as a Petri-net transition has a fixed multi-set of {\em
pre-places}, each transition in the Join Calculus has a fixed {\em join pattern}
defining its input signals.  The key difference is that the Join Calculus is
higher-order, allowing signals to be passed as values, and for the output
signals to depend on its inputs---unlike Petri-nets where the {\em post-places}
of a transition are fixed.  This simple modification allows use of continuations
to support functions.  Moreover, while nets are static at runtime, a Join
Calculus program can create new instances of definitions (containing signals and
transition rules) at runtime, and although these cannot match on existing
signals, existing transitions can send messages to the new signals.

\section{Mapping Programs to Heterogeneous Hardware}\label{sec:construction}
We will use the same simple hardware model as in our previous work. This
considers each processor to be closely tied to a local memory. It then defines
interconnects between these. The construction itself will be concerned with a
finite set of {\em processors} $P$, a set of directed {\em interconnects}
between these $I \subseteq P \times P$, and a {\em computability} relation $C
\subseteq P \times R$ (where $R$ is the set of transition rules in the program),
such that $(p,r) \in C$ implies that the rule $r$ can be executed on the
processor $p$.  In our example, we take $P = \{\mathtt{x}, \mathtt{y}\}$, $I =
\{(\mathtt{x}, \mathtt{y}), (\mathtt{y}, \mathtt{x})\}$ and a computability
relation equal to $P \times R$.
However, it is easy to imagine more complex scenarios---for instance, if one processor lacked floating point support, $C$ would not relate it to any transitions using floating point operations.

A scheduler will also need a cost model, however this is not needed for this
work. We would expect an affine (i.e. latency plus bandwidth) cost for the
interconnect.
In practice, this and the approximate cost of each transition on each processor would be given by profiling information.

There are two parts to our construction. Firstly, we produce a copy of the
program for each $p \in P$, omitting any transition rules $r$ for which $(p,r)
\not\in C$, giving box ``B'' of Figure~\ref{fig:code}.

Secondly, we add transitions that correspond to possible data transfers (box ``C'').
This requires one rule per signal and interconnect pair. However, the higher-order
nature of the Join Calculus means these need more careful definition than in our
Petri-net work to preserve locality. Specifically, when a signal value such as
\texttt{k} is transferred it needs to be modified so that it becomes local to
the destination processor. This maintains the invariant that the
`computation transitions' introduced by the first part of the construction can
only call signals on the same processor.

\section{Workers in place of Resource Constraints}\label{sec:workers}
In the Petri-net version of this work, there was a third part to the
construction. We introduced {\em resource constraint places} to ensure that
each processor or interconnect only performed one transition at once.
Equivalent signals would be illegal in the Join Calculus, as they would need to
be matched on by transitions from multiple definition instances (since processor
time is shared between these). Changing the calculus to allow this would make it
harder to generate an efficient implementation. Instead, we introduce the
notion of {\em workers} to the semantics.

Rather than allowing any number of transition firings to be mid-execution at
a given time, we restrict each worker to performing zero or one firing at a
time. We also tag each transition with the worker that may fire it. In our
example, we would have four workers: \texttt{x}, \texttt{y}, \texttt{(x,y)} and
\texttt{(y,x)}. The \texttt{\_x} and \texttt{\_y} copies of the original program
are tagged with the \texttt{x} and \texttt{y} CPU workers respectively, while the data transfer
transitions are tagged with the relevant interconnect worker.

To accommodate vector processors such as GPUs, we augment $n$ copies of an existing transition with a single merged transition.
The new transition will take significantly less time than performing the $n$ transitions individually.
Obviously, a real implementation will not enumerate these merged transitions, but we can view it this way in the abstract.
A similar argument also applies to data transfers, where we can benefit from doing bulk operations.

This gives the formal semantics for our calculus as defined in Figure~\ref{fig:semantics}.
We give this for an abstract machine, however just as Java trivially compiles to the JVM, our non-nested language can trivially be compiled to this JCAM. 
We also use a small step semantics rather than the ChAM \cite{Fournet1996} or rewriting \cite{Odersky2000} style used previously, as this is more appropriate for our ongoing work on analysis and optimisation.
Each of the workers can be either processing a transition rule, or \texttt{IDLE}.
The initial state is for all workers to be \texttt{IDLE}, and some messages corresponding to program arguments to be available in $\Gamma$.
\begin{figure}
  \small
  \textbf{Domains}:
  \begin{align*}
    (f,t), (f, \theta) \in \mathrm{SignalValue} &= \mathrm{Signal} \times \mathrm{Time}\\
    \Gamma \in \mathrm{Environment} &= \textbf{m}(\mathrm{SignalValue} \times \mathrm{Value}^{*}) & \text{(messages available)}\\
    t \in \mathrm{Time} &= \mathbb{N}_0\\
    \Sigma \in \mathrm{GlobalState} &= \mathrm{Worker} \rightarrow (\mathrm{LocalState} \cup \{\mathtt{IDLE}\})\\
    (l, \theta, \sigma) \in \mathrm{LocalState} &= \mathrm{Label} \times \mathrm{Time} \times \mathrm{Value}^{*} & \text{(program counter, context, local stack)}\\
    v \in \mathrm{Value} &= \mathrm{SignalValue} \cup \mathrm{Primitive}
  \end{align*}
  
  \textbf{Rules} (judgement form of $\Gamma, t, \Sigma \rightarrow \Gamma', t', \Sigma'$):
  \begin{align*}
    \Gamma + \Delta,\; t,\; \Sigma + \{w \mapsto \mathtt{IDLE}\} &\rightarrow \Gamma,\; t,\; \Sigma + \{w \mapsto (l_0, \theta, \vec{v_1} \cdot \ldots \cdot \vec{v_n})\} & \text{(fire)}\\
    \shortintertext{where  $\Delta = \{((f_1, \theta), \vec{v_1}), \ldots, ((f_n, \theta), \vec{v_n})\}$ and $r^w = f_1(\ldots) \& \ldots \& f_n(\ldots) \{ l_0, \ldots \}$}
    \Gamma,\; t,\; \Sigma + \{w \mapsto (\texttt{EMIT}^l, \theta, \vec{v} \cdot s \cdot \sigma)\} &\rightarrow \Gamma + (s, \vec{v}),\; t,\; \Sigma + \{w \mapsto (\mathrm{next}(l), \theta, \sigma)\} & \text{(emit)}\\
    \Gamma,\; t,\; \Sigma + \{w \mapsto (\texttt{CONSTRUCT<}f\texttt{>}^l, \theta, \vec{v} \cdot \sigma)\} &\rightarrow \Gamma + ((f,t), \vec{v}),\; t+1,\; \Sigma+\{w \mapsto (\mathrm{next}(l), \theta, \sigma)\} & \text{(construct)}\\
    \Gamma,\; t,\; \Sigma + \{w \mapsto (\texttt{LOAD.SIGNAL<}f\texttt{>}^l, \theta, \sigma)\} & \rightarrow \Gamma,\; t,\; \Sigma + \{w \mapsto (\mathrm{next}(l), \theta, (f,\theta) \cdot \sigma)\} & \text{(load)}\\
    \Gamma,\; t,\; \Sigma + \{w \mapsto (\texttt{FINISH}^l, \theta, \sigma)\} &\rightarrow \Gamma,\; t,\; \Sigma+\{w \mapsto \mathtt{IDLE}\} & \text{(finish)}
  \end{align*}
  \caption{Non-Nested Join Calculus Abstract Machine (JCAM) Semantics}\label{fig:semantics}
\end{figure}
Unmapped programs can be considered to have just a single worker.

\section{Future Work on Scheduling}\label{sec:scheduling}
As before, we rely on a scheduler to be able to make non-deterministic
choices corresponding to the fastest execution---and clearly these need to be
made quickly. In this section, we briefly discuss our thoughts on this problem.

It is clear that to optimise the expected execution time, we need transition
costs, and also a probability distribution for the output signals of a
transition. We believe that these could be effectively provided by profiling.
This is already commonly used in auto-tuning (i.e. transition costs) and branch
predictors (i.e. signal probabilities). It could also be used for {\em
ahead-of-time scheduling} or just for determining baselines.

For practical scheduling, it is most likely that a form of machine learning will be used to adapt to new architectures.
This has been used successfully for streaming applications \cite{Wang2010}, which are not dissimilar to a very restricted Join Calculus.
Existing implementations of the Join Calculus have not considered the
scheduling problem, and simply pick the first transition found to match. In
order to maintain this simplicity, we would consider whether the output of such
a learning algorithm could be a priority list of transitions, that evolves over time to offer some load balancing.

Prior work has shown it is best to check for transition firings each time a
message is emitted, rather than having a separate firing process
\cite{Turon2011}.  This will result in a queue of transitions (perhaps picked
by a first-found approach). For load balancing, idle workers could then {\em steal}
from these transition queues.  However, unlike in standard {\em work stealing},
there are two possible forms of stealing---as well as taking a matched transition, individual messages could be taken by first decomposing some of the existing matches.

\section{Conclusions}\label{sec:conclusion}
In this paper, we have adapted our existing work on mapping Petri-net programs
to heterogeneous architectures to the Join Calculus.  In doing so, we showed how
to remove the problematic environments introduced by nested definitions, and
also avoid global matching on resource signals by modifying the semantics slightly to incorporate {\em workers}.
This allows programs to be agnostic of the architecture they will run on, with any placement and scheduling choices that depend on the architecture being left in the program.
We have also listed several challenges for building a scheduler that can optimise over these choices, and initial ideas to solve them.
Such an implementation is our current research goal.

\subsection*{Acknowledgements}
We thank the Schiff Foundation, University of Cambridge, for funding this
research through a PhD studentship; the anonymous reviewers for useful comments
and corrections; and also Tomas Petricek for useful discussions about these
ideas.


\bibliographystyle{plain}

\end{document}